\documentstyle{mn}

\def\today{\ifcase\month\or
 January\or February\or March\or April\or May\or June\or
 July\or August\or September\or October\or November\or
 December\fi\space\number\day, \number\year}

\def\todmy{\number\day\space\ifcase\month\or
 January\or February\or March\or April\or May\or June\or
 July\or August\or September\or October\or November\or
 December\fi\space\number\year}

\newcommand{\bdisp} {\begin{displaymath}}
\newcommand{\edisp} {\end{displaymath}}
\newcommand{\beqn} {\begin{equation}}
\newcommand{\eeqn} {\end{equation}}
\newcommand{\beqr} {\begin{array}}
\newcommand{\eeqr} {\end{array}}

\title{The effect of a cusped matter distribution on the formation 
of brightest cluster members}
\author[E. Athanassoula, A. Garijo and  C. Garc\'{\i}a
G\'omez]{E. Athanassoula$^{1}$, A. Garijo$^{2}$  and C. Garc\'{\i}a
G\'omez$^{2}$ \\
$^{1}$ Observatoire de Marseille,
2 place Le Verrier,
F-13248 Marseille Cedex 4, France\\ 
$^{2}$ Dep. Enginyeria Inform\`atica i Matem\`atiques,
Escola T\`ecnica Superior d'Enginyeria,\\
Universitat Rovira i Virgili,
43006 Tarragona, Spain}

\date{Accepted .
      Received ;
      }

\pagerange{\pageref{firstpage}--\pageref{lastpage}}
\pubyear{1999}

\begin{document}

\maketitle

\label{firstpage} 
\begin{abstract}

We present N-body simulations of galaxy groups embedded in a common halo
of matter. We study the influence of the different initial conditions upon the
evolution of the group and show that denser configurations evolve
faster, as expected. We then concentrate on the influence of the initial 
radial density profile of the common halo and of the galaxy
distribution. We select two
kinds of density distributions, a singular profile (modeled by a
Hernquist distribution) and a profile with a flat core (modeled by a Plummer
sphere). In all cases we witness the formation of a central massive
object due to mergings of individual galaxies and to accretion of
stripped material, but both its formation history and its properties
depend heavily 
on the initial distribution. In Hernquist models the formation is due
to a ``burst'' of mergings in the inner parts, due to the large initial
concentration of galaxies in the center. The merging rate is much
slower in the initial phases of the evolution of a Plummer
distribution, where the contribution of 
accretion to the formation of the central object is much more
important. The central objects 
formed within Plummer distributions have projected density 
profiles which are not in agreement with the radial profiles of
observed brightest cluster members, unless the
percentage of mass in the common halo is small.
On the contrary the central object formed in initially cusped models
has projected radial profiles in very good agreement with those of
brightest cluster members, sometimes also showing luminosity
excess over the $r^{1/4}$ law in the outer parts, as is observed in cD
galaxies.
 
\end{abstract}


\section{Introduction}

\indent
Brightest cluster members (BCMs) are giant ellipticals found in the
central parts of clusters of galaxies. Some of these galaxies are
further classified as cD galaxies because they show in their external
parts a projected surface brightness excess over the standard $r^{1/4}$ law.
Several models have been proposed for the origin of BCMs in general
and of cDs in particular. A first class of models relies on the presence
of cooling flows in clusters of galaxies. If the central cluster
density is sufficiently high, 
intra-cluster gas can gradually condense and form stars at the bottom of
the potential well (Cowie \& Binney 1977; Fabian \& Nulsen 1977). Such models,
however, imply the formation of a lot of new stars, of which there is
no good observational evidence. A second class of models relies on the
dynamical mechanisms of merging, cannibalism and
accretion (Ostriker and Tremaine 1975; Ostriker
and Hausman 1977; Hausman and Ostriker 1978; McGlynn and Ostriker
1980 etc.). Schematically the central giant  
galaxy is formed by the merger of smaller galaxies and its growth
continues by further mergings and cannibalism. It also
accretes material stripped from the outer parts of the other galaxies, by
interactions between galaxies, or between a galaxy and the common
background. It has been proposed (Gallagher \& Ostriker 1972; Richstone 
1976) that such material could be responsible for the luminous envelope of cD 
galaxies. Merritt (1984) argued that most of the formation happens during 
cluster collapse. 

Most of the mass of clusters or groups of 
galaxies is not bound to the galaxies that constitute them. Instead it is 
distributed in a common background encompassing the whole cluster and
it includes 
the high temperature gas emitting X-rays and an unknown fraction of
dark matter. 
This unknown mass distribution must have influenced the formation mechanism of
BCMs. Nevertheless, considerable uncertainty still remains
as to what fraction of the total cluster mass resides in the dark matter, and 
how it is distributed. 

Observations of galaxy clusters can be used to constrain their mass
distribution.
The first attempts to obtain the profile of this mass distribution
were based on galaxy counts and the projected velocity
dispersion of the galaxy system. They rely on the assumption that the
cluster has reached virial equilibrium and depend on what is
assumed about the possible anisotropy of the system. They lead to mass
distributions possessing central cores where the mass density is nearly 
constant (e.g. Kent \& Gunn 1982). More recent studies, however, point
to a mass 
distribution with a cusp in the central parts (Carlberg et al. 1997).
 
For some galaxy clusters which are very luminous in X-rays, many authors have 
studied the total mass distribution under the assumption that the hot gas is in
hydrostatic equilibrium within the total potential. The
correlation between the velocity dispersion of the galaxies and the X-ray 
temperature (e.g. Wu et al. 1998), as well as  the observational evidence
that the dynamical properties of clusters have evolved little since
$z\sim 0.8$ (Mushotzky \& Scharf 1997;  
Bahcall, Fan \& Chen 1997; Henry 1997; Rosati et al. 1998; Vikhlinin
et al. 1998, 
etc.) indicate that clusters are globally relaxed structures. Further
supportive evidence comes from the fact that Crimele et al. (1997) find
a relation between the gas and galaxy densities consistent with
the assumption of hydrodynamical equilibrium. Thus the  
Jeans equation together with the hypothesis of hydrostatic equilibrium
can be used to describe clusters globally, 
except for clusters which are still in the process of
formation, or of merging with another cluster. In most cases X-ray mass
estimates have been obtained with models with flat cores. More recently, however,
Tamura et al. (2000) have shown that models of the total mass distribution with
cusps give better fits to the data of the cluster A1060 than models with flat
cores. 

Yet a third way to derive the mass in a cluster
is to use the gravitational lensing effect which is produced when the
images of background galaxies are distorted by the mass concentration of a
galaxy cluster. In the case of strong lensing background galaxies are distorted
as giant blue arcs. A massive 
foreground cluster can also change the shapes (shear effect) and the number 
density (magnification) of the faint background population. These
effects are 
known as weak lensing. Arcs have been used intensively to map the matter
distribution in the central parts of clusters (Fort \& Mellier 1994; Mellier
1999 and references therein). In particular considerable
progress was achieved from recent spectacular images of arc(let)s obtained with
the HST. Many different mass distributions, including profiles with cores or
cusps, have been used in the modelling. There seems to be agreement that, if
cores exist, their radius should be less than $50h_{100}^{-1}$ Kpc
(e.g. Mellier 1999 and references therein). Except for that, no clear consensus
seem to have been reached on the form of the radial profile in the innermost
parts.

Comparison of the results obtained with the various methods described so far
lead to considerable controversy. In particular, results from strong lensing are
a factor $2-4$ higher that results from X-rays. (e.g. Loeb
\& Mao 1994, Miralda-Escud\'e \& Babul 1995, Tyson \& Fischer
1995; Wu et al. 1998, Mellier 1999). This could be explained by the fact that,
although the X-ray gas should be globally in gravitational equilibrium,
locally in the center-most parts virial equilibrium may not be
achieved. It should, however, be noted that models with smaller core radii, with
cusps, or more generally with higher masses in the center-most parts, reduce the
differences between the results obtained with the different methods (e.g. Wu et
al. 1998, Makino \& Asano 1999, Markevitch et al. 1999).

N-body numerical simulations of
hierarchical gravitational collapse of an ensemble of cold collisionless
particles lead to halo radial profiles with a cusp in the central parts.
Navarro, Frenk and White (1996) argued that such profiles can be
fitted by 
a ``universal'' profile of the form:
\begin{equation}
\rho(r) = \frac{\rho_s}{(r/r_s)(1+r/r_s)^2}
\end{equation}
where $\rho_s$ and $r_s$ are the characteristic density and length
respectively.  
This profile was first obtained in the context of the standard cold dark matter
cosmology, but subsequent numerical studies have shown that this profile is
independent of halo mass, of the initial density fluctuation and of cosmology
(e.g. Cole \& 
Lacey 1996; Navarro Frenk and White 1997; Eke, Navarro \& Frenk 1998;
Jing 1999). Simulations with higher numerical resolution 
result in profiles which are steeper in the central parts, i.e. of the
form $\rho 
\sim r^{-1.4}-r^{-1.5}$ (Fukushige \& Makino 1997, Moore et al. 1998), while 
cosmologies using collisional dark matter point to even steeper 
($\rho \sim r^{-2}$) profiles (Yoshida et al. 2000;
Moore et al. 2000), at least in the limit of short mean free paths.

For these considerations it seems appropriate to compare the evolution of 
clusters of galaxies within backgrounds with different mass profiles, in hope
that this may lead to clues about which kind of mass distribution is more
realistic. Several N-body simulations of the dynamical evolution of
large groups or 
clusters of galaxies with each galaxy being represented by many
particles have been reported in the literature.  Funato, Makino \&
Ebisuzaki (1993), Sensui, Funato \& Makino (1999) and Garijo, Athanassoula
\& Garc\'{\i}a-G\'omez (1997, hereafter GAG) use model clusters where
initially all the mass is in the galaxies.  
Funato, Makino \& Ebisuzaki (1993) argue that the Faber-Jackson
relation is a result of the evolution of galaxies driven by
interaction with other galaxies and with the tidal field of the parent
cluster. Sensui, Funato \& Makino (1999) follow the evolution of
a cluster of 128 galaxies initially in virial equilibrium with all mass
being attached to the galaxies. They find that within a few cluster
crossing times half of the total mass has escaped individual galaxies,
and that the amount of stripped material is larger for galaxies in the
central regions. The density profile of the common halo thus formed is
cuspy, with a radial dependence roughly $\rho \sim r^{-1}$ in the
central region and $\rho \sim r^{-4}$ in the outer regions. Bode et al. (1994), GAG and
Dubinski (1998) witnessed in their simulations the formation of a
massive central galaxy. Bode et al. (1994) modelled poor clusters with
50 galaxies and a varying percentage of the mass in a common
background. They find multiple nuclei in the central object in between
10 and 40 per cent of the time. Increasing the percentage of mass in
the common background slows the merging rate and this can be
sufficient for stalling the formation of the central object if 90 per
cent of the mass is in a common background. GAG studied in detail the
properties of the central object and found good agreement with those of
brightest cluster members. The object formed is roughly spherical,
oblate or mildly triaxial. In the latter cases the orientation of the
central object correlates well with that of the central group. The
triaxiality is in general stronger in the outer parts of the central
object. They also find three types of projected density
distributions. Objects in the first category have profiles well fitted
by an $r^{1/4}$ law. In the second category of objects the radial
profile follows the $r^{1/4}$ law only in the main part of the galaxy
and falls bellow it in the outer parts. In the third, most interesting,
category the projected surface density of the objects follows the
$r^{1/4}$ law in the main body of the galaxy, but is systematically
above it in the outer parts, as is the case for cD galaxies. The
projected velocity dispersion profiles of the central objects also
agree with those of brightest cluster galaxies. Dubinski
(1998) used an N-body simulation of a cluster of galaxies in a
hierarchical cosmological model. He replaced 100 dark matter halos at
$z$ = 2 with self-consistent disc+bulge+halo galaxy models, thus
attaining high resolution. He witnessed formation of the central
galaxy through the merging of several galaxies along a filament. The
central object was flattened and triaxial, showing alignment with the
primordial filament. Its projected surface density was well described by an
$r^{1/4}$ law and showed no excess in the outer parts, as one would
expect from a cD. The kinematics of the object also was in good
agreement with the observations.

In this paper we will study the influence of the dark matter
distribution on the evolution of the cluster and on the formation and
properties of the central galaxy. Our aim is 
to show that the observable properties of this galaxy can be used to constrain
the dark matter distribution in the central parts. The paper is organized as
follows. In section~2 we present the simulations and the initial conditions. 
In section~3 we describe the different evolution of the
various models and in section~4 we show how the different initial conditions 
influence the formation process and the properties of the central 
giant galaxy. Section~5 discusses the main assumptions and
simplifications entering in our simulations. Our results are
discussed and summarised in Section~6. 
 
\section{Simulations}
\label{sec:simulations}

We have performed a total of $17$ simulations following the time
evolution of a group of $50$ galaxies in virial equilibrium within a common
halo. In nearly all cases we used a total of $100\,000$ particles with the same
mass. A fraction of
these particles was used to represent the galaxies and the rest were
used for the common halo. In one simulation only one third of the total mass was
in galaxies and for this case we used $200\,000$ particles, while maintaining 
the total mass constant, in order to
have galaxies represented by a reasonable number of particles.

The common halo was
represented alternatively as a Plummer or as a Hernquist sphere
(Hernquist 1990).  
The density distribution of a Plummer sphere follows the law:
\begin{equation}
\rho_p(r) = \frac{3M_{ch}}{4\pi a^3_{p}}\left(\frac{a^2_{p}}
{r^2+a^2_{p}}\right)^{5/2}.
\end{equation}
This distribution is characterised by the parameters $M_{ch}$, which represents 
the total mass in the common halo, and $a_p$ which represents the scale-length of 
the distribution. The Hernquist distribution follows the law:

\begin{equation}
\rho_h(r) = \frac{M_{ch}}{2\pi}\frac{a_h}{r}\frac{1}{(r+a_h)^3}.
\end{equation}
Two parameters are also necessary to determine this distribution: the total 
mass in the common halo $M_{ch}$ and the scale-length $a_h$. Note that while 
the Plummer distribution has a central density of 
${3M_{ch}}/{4\pi a_p^3}$, the Hernquist 
distribution is singular at the origin. Its projected surface density
follows closely the $r^{1/4}$ law for a large fraction of the radius 
(Hernquist 1990). It gives a cusped density distribution in the central
parts which falls as $\rho(r) \sim r^{-1}$ for small radii and as
$\rho(r) \sim r^{-4}$ at large radii. 

For a given mass of the common halo $M_{ch}$, the parameters $a_p$ and 
$a_h$ were selected as to give a common halo with the same total energy. This
 gives the following relation between the two parameters:

\begin{equation}
a_h = \frac{16}{9\pi}a_p.
\end{equation}

\noindent In this way we can compare simulations with the same global 
properties changing only the mass profile of the common halo.

\begin{table}
\centering
\caption{Initial conditions}
\begin{tabular}{llllll}\\ 
\multicolumn{6}{c}{Plummer models}\\\hline
Name & $a_p$ & $f$ & $r$ & $N_{pg}$ & Common Halo\\\hline
P5f3/2r1 & 5 & 3/2 & 1 & 1000 & Live \\
P10f3/4r1 & 10 & 3/4 & 1 & 1000 & Live \\
P15f3/2r1 & 15 & 3/2 & 1 & 1000 & Live \\
P10f3/2r1 & 10 & 3/2 & 1 & 1000 & Live \\
P10f1r1 & 10 & 1 & 1 & 1000 & Live \\
P10f3/4r1 & 10 & 3/4 & 1 & 1000 & Rigid \\
P10f1r2 & 10 & 1 & 2 & 1333 & Live \\
P10f1r4 & 10 & 1 & 4 & 1600 & Live \\
P10f1r6 & 10 & 1 & 6 & 1714 & Live \\
P10f1r10 & 10 & 1 & 10 & 1818 & Live \\
P10f1r20 & 10 & 1 & 20 & 1904 & Live \\ 
P10r1 & 10 & $\infty$ & 1 & 1000 & Live \\
P10r20 & 10 & $\infty$ & 20 & 1904 & Live \\\hline
\multicolumn{6}{c}{}\\
\multicolumn{6}{c}{Hernquist models}\\\hline
Name & $a_h$ & $f$ & $r$ & $N_{pg}$ & Common Halo\\\hline
H10f1r1/2 & 5.65 & 1 & 0.5 & 1333 & Live \\
H10f1r1 & 5.65 & 1 & 1 & 1000  & Live \\
H10f1r5 & 5.65 & 1 & 5 & 1666 & Live \\
H10f1r20 & 5.65 & 1 & 20 & 1904 & Live \\\hline
\end{tabular}
\end{table}

The positions and velocities of the particles in the common halo are
drawn according to the chosen Plummer (cf. eg. Aarseth, Henon \&
Wielen 1974) or Hernquist model (Hernquist 1990).
The positions and velocities of the center of masses of the galaxies
are drawn from the same distribution as the common halo,
except that we impose a cut-off radius which is a fraction of the half
mass radius $R_{1/2}$ of  
the common halo. Thus the galaxies are distributed within a radius selected as
\begin{equation}
R_{cl} = f R_{1/2},
\end{equation}
\noindent where $f$ is a parameter which is used to determine the initial 
concentration of the galaxy system. Finally, we also vary the 
ratio between the mass in galaxies and the mass in the common halo. 
This mass ratio is given by the parameter:
\begin{equation}
r = \frac{N_g M_g}{M_{ch}},
\end{equation}
\noindent where $N_g$ represents the initial number of galaxies and
$M_g$ the initial mass of each galaxy.
We use in all cases $N_g=50$ identical galaxies, whose density profiles 
follow a Plummer law with a scale-length of $a_{gal}=0.2$. The mass of each 
galaxy, $M_g$, will depend on the mass ratio selected for each
particular initial condition, i.e. on the value of $r$.

We repeated one of the simulations using a 
rigid halo which is not modified during the whole simulation, 
instead of a live halo composed of particles. We did this in order to 
study the effect of dynamical friction and the possible differences
introduced by a halo capable of responding to  
the variation of the mass distribution as the central giant galaxy is formed. 

In Table~1 we give more detailed information
about our simulations. In {\it Column}~1 we list the simulation name, coded
conveniently so as to describe the initial conditions of each
simulation. The first 
letter P or H is used to note the type of mass 
distribution,  
P for Plummer models and H for Hernquist models. This letter is 
followed by one or two digits indicating the scale-length of the common halo in
simula\-tion units. For the case of Hernquist models we indicate the
scale-length  
of the Plummer model with the same total energy. Thus from the name of the 
simulations we can readily see which of the different simulations are directly 
comparable. After these digits comes the letter $f$ followed by
the value  
of this parameter and finally the letter $r$ followed by the value of
the mass ratio. 
In this work we have considered values of $r$ between $1/2$
and $20$,  
thus covering a range of more than one order of magnitude. The three
values of the parameters  
determining the initial conditions are repeated for clarity in {\it
Columns}~2,  
3 and 4. In {\it Column} 2 for the case of the Hernquist simulations we list
the value  
of the scale-length of the Hernquist distribution. In {\it Column}~5
we indicate the  
total number of
particles used to represent each galaxy and
finally in {\it Column}~6 we indicate if the common halo is rigid or a live 
one evolving freely. In two of the simulations, P10r1 and P10r20,
the initial positions of the galaxies were 
selected anywhere within the common mass distribution without any cut-off. 
Finally in the 
Hernquist simulation with low galaxy-to-common-halo mass ratio, namely 
H10f1r1/2, the total 
number of particles was doubled, while the total mass was kept constant.

To evolve these systems we used the GRAPE-3AF system of the Marseille
Observatory using a version of 
a treecode specially adapted for this system (Athanassoula et
al. 1998). This version, since the tree is descended for a whole
group of particles, rather than for every particle separately, is more
precise than a standard treecode (Barnes \& Hut 1986).
For the opening angle we used $\theta = 0.7$, for 
the time step $\delta t = 0.0078125$ and for the softening parameter
$\epsilon = 0.0625$. With these values the total energy was 
conserved to within a few parts in a thousand. The simulations with a
Plummer matter distribution were 
followed for a total of roughly $60$ computer units or, equivalently, roughly
$7680$ time steps.
Each simulation required about $24$ hours of CPU time on the Marseille five
board GRAPE-3AF system. The evolution of simulations with a Hernquist
distribution was followed for twice as long, to follow best the evolution
of the central object.

In this paper, we use the same units as in GAG,
where the simulation units
were selected as $G = 1$, the total mass $M_T = 50$ and the typical
scale-length  
of a galaxy $a_{gal} = 0.2$. With the aim of comparing with the
observations we  
set $M_T = 1.5 \times 10^{13} M_\odot$ and 
$a_{gal}$ = 6 Kpc. These 
values give a total of $1.4 \times 10^8$ years for the time
unit, $210$ Km/s for the velocity unit and a total simulation time of 
$T = 8.4$ and 16.8 Gyr, for Plummer and Hernquist distributions
respectively. Note that this selection of conversion factors, albeit
reasonable, is not unique and neighbouring  values  
could also be used. This should be taken into account when comparing our 
simulations with observational data, hence agreements to within a factor of two
should be considered to be quite satisfactory.

\section{Evolution of the simulations}
\label{sec:evol}

\subsection{A general impression from snapshots}
\label{snapshots}

\begin{figure*}
\includegraphics{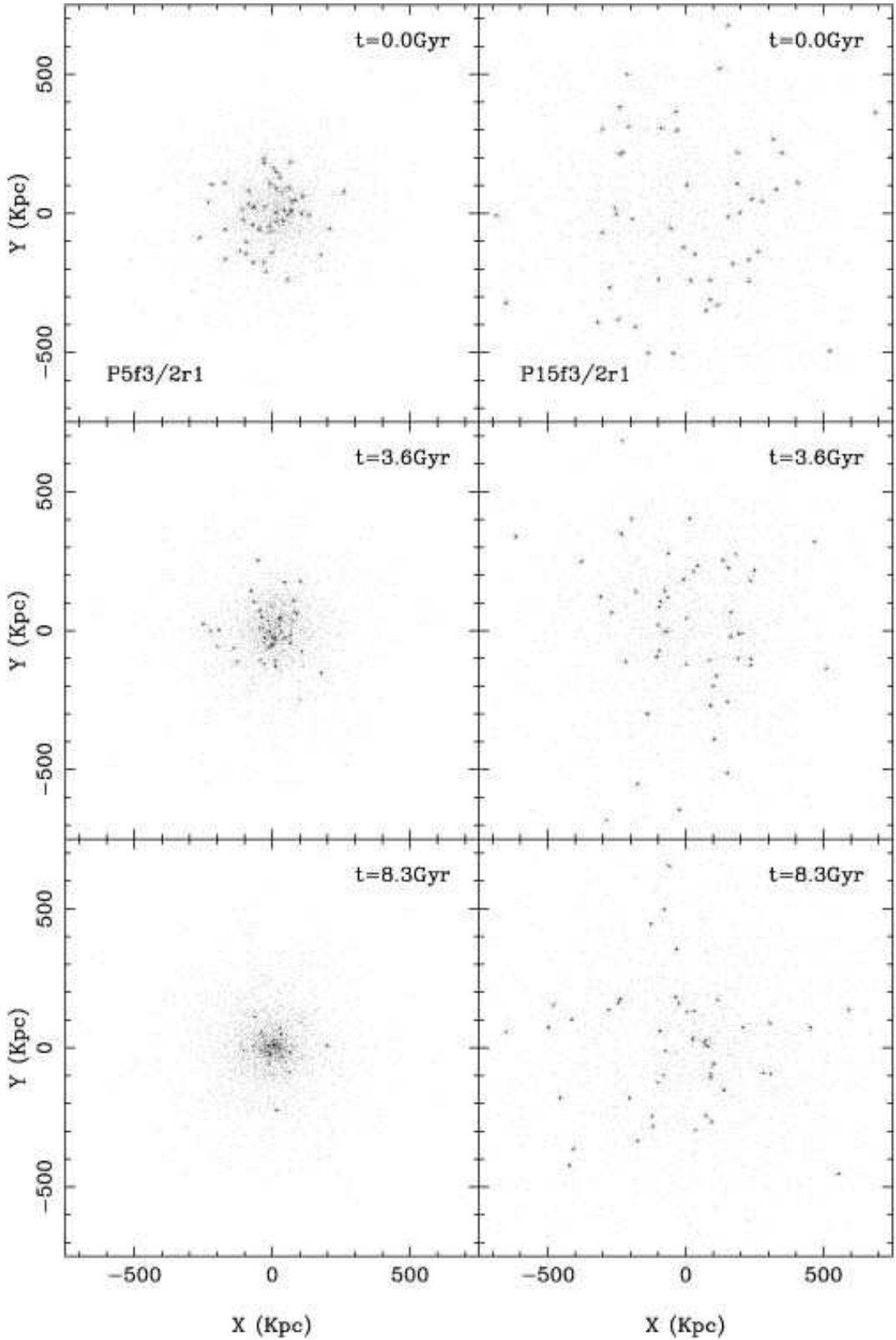}
\vspace{22cm}
\caption{Comparison of the evolution of two simulations, one
corresponding to a small scale-length of the common halo and galaxy
distribution (left panels) and the
other to a large one (right panels). Three times are shown for each
simulation, the initial one, one not far from half way through the
simulation and the
third one at the end. The times are given in the upper right corner of
each panel and the simulation name in the top panel of each column. 
Only one particle in 20 is plotted.}
\label{evol1}
\end{figure*}

\begin{figure*}
\includegraphics{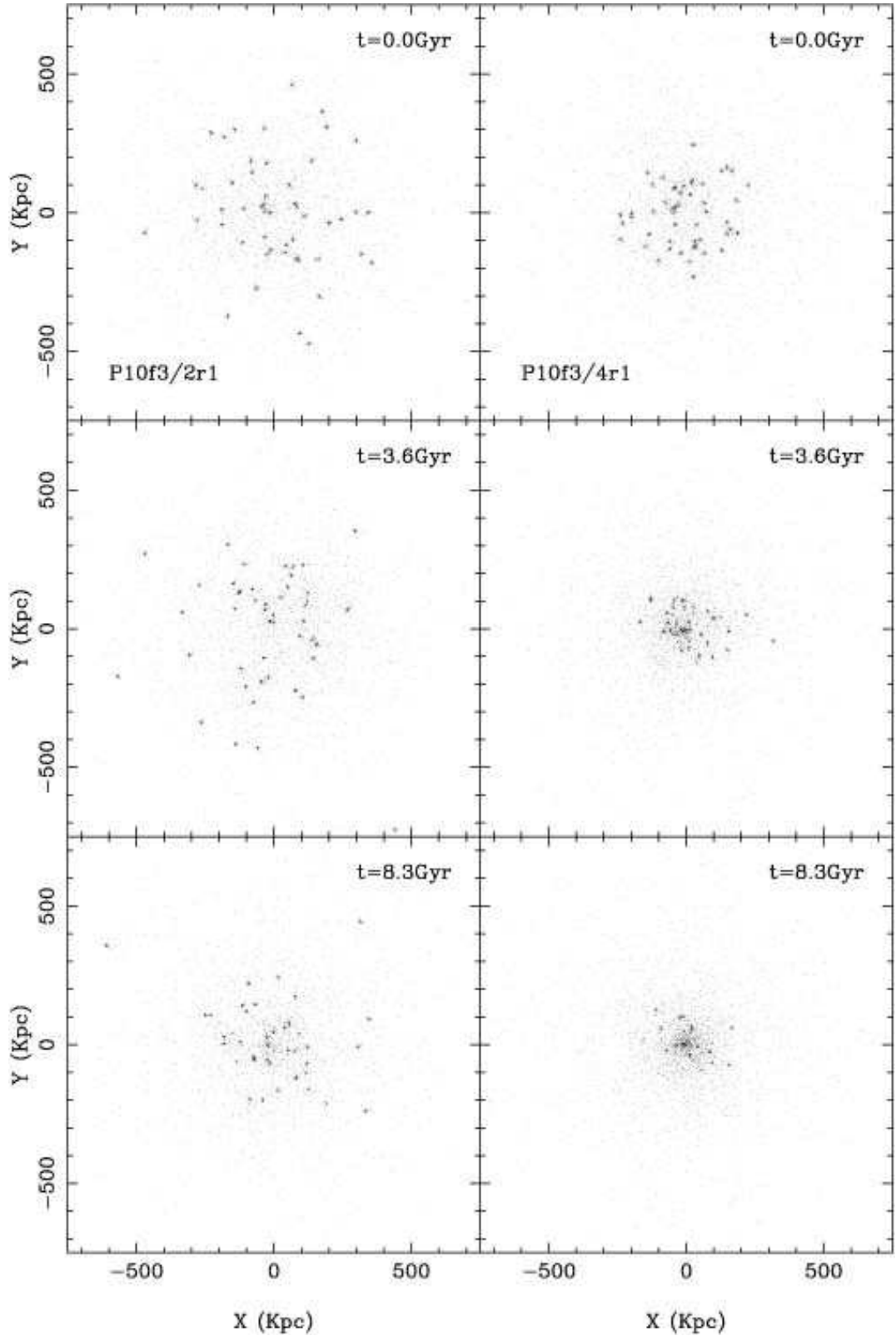}
\vspace{22cm}
\caption{Effect of the initial cut-off radius of the galaxy
distribution 
on the evolution of the group. If the cut-off radius is smaller,
i.e. if the galaxy system is initially more
dense (right panels) then the evolution of the system is faster. The layout of
this figure follows that of Fig.~\ref{evol1}.}
\label{evol2}
\end{figure*}

\begin{figure*}
\includegraphics{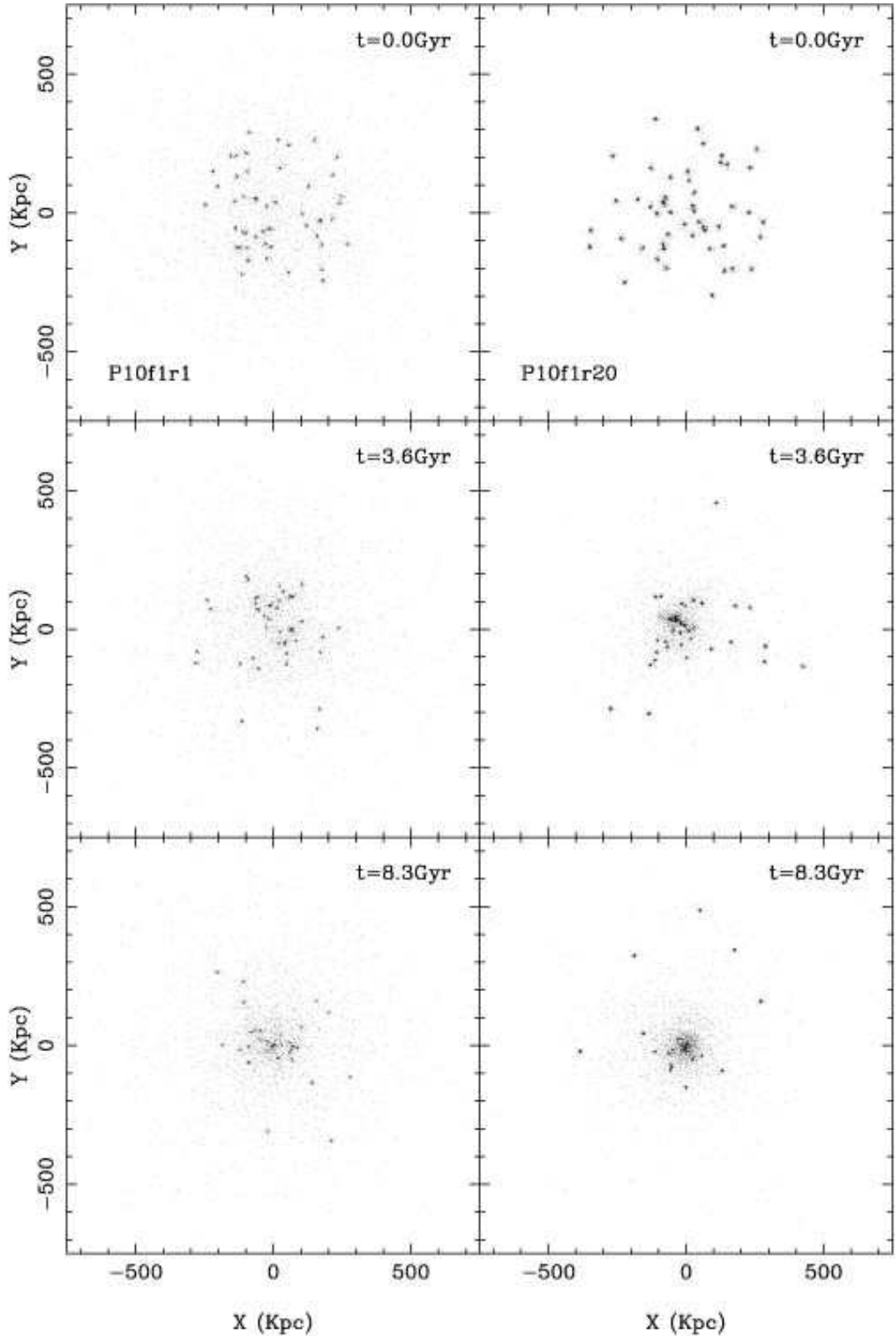}
\vspace{22cm}
\caption{Effect of the ratio of the mass in galaxies to the mass in the common
halo on the evolution of the group. If the mass is initially placed
mainly in the galaxies 
(right panels) the system evolves faster than if a larger fraction of
the mass is in the common halo (left panels). The layout of the figure
follows that of Fig.~\ref{evol1}.}
\label{evol3}
\end{figure*}

\begin{figure*}
\includegraphics{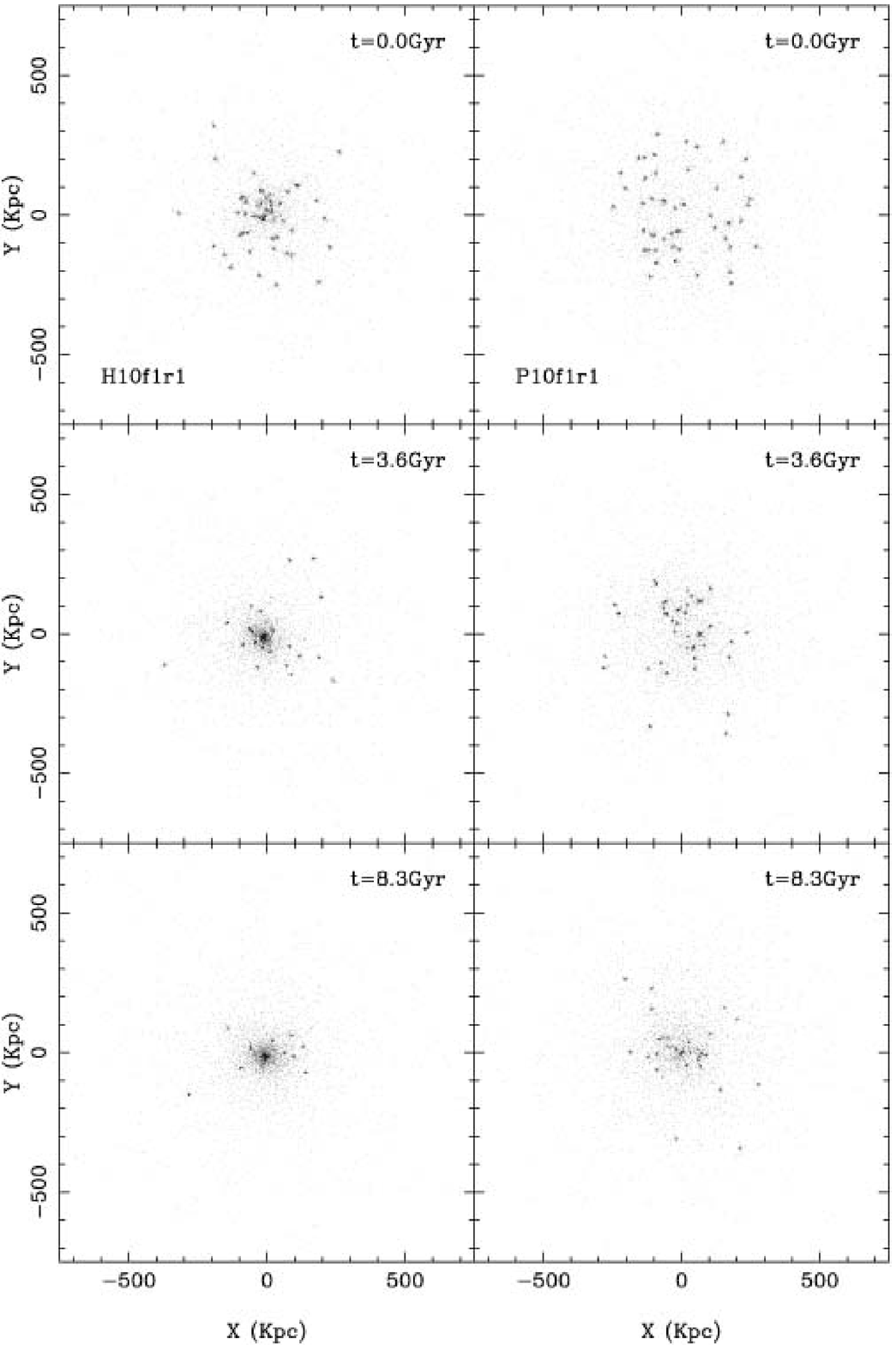}
\vspace{22cm}
\caption{Effect of the different common halo and galaxy mass distribution. The
left
panels show the evolution of the galaxy system within a Hernquist model
while the right panels shown the evolution of the galaxies within a
Plummer model of the same total energy. The central galaxy grows faster
in the Hernquist models. The layout of the figure
follows that of Fig.~\ref{evol1}.}
\label{evol4}
\end{figure*}

As in the simulations without common halo (GAG) the most noticeable 
result of the dynamical
evolution is the formation of a giant central galaxy in the central
parts of the group. The rate of formation of this galaxy, as well as its 
final extension, depend on the properties 
of the common halo and of the galaxy system. This can be seen qualitatively
in Figs.~\ref{evol1} to \ref{evol4}, where we 
show respectively the influence of the scale-length, of the initial 
extension of 
the galaxy distribution, of the ratio of the mass in galaxies 
to the mass in the common halo and the influence of different mass profiles. 

Amongst our simulations with Plummer background  the one that starts
more centrally concentrated  
is P5f3/2r1 and the one that starts off with the least central 
concentration is P15f3/2r1. Since the other parameters are the same in 
the two simulations, the comparison of the two evolutions, given in 
Fig.~\ref{evol1}, shows the effect of the scale-length
of the galaxy distribution and of the common halo, which
changes by a factor of three between the two cases. The more
concentrated systems, with shorter scale-length, have faster 
evolution rates. In this way in simulation P5f3/2r1 a central galaxy is
rapidly formed which controls thereafter the dynamical evolution of the system,
merging with the satellite galaxies in a shorter time than in the rest of the
simulations. On the other hand, in the simulation P15f3/2r1 no central
giant galaxy is formed and practically all the galaxies survive until the end of
the simulation. 

In Fig.~\ref{evol2} we compare the evolution of simulations P10f3/2r1 and
P10f3/4r1 to understand the effect of changing the initial
cut-off radius of the galaxy distribution. Since the number of
galaxies is initially the same in both cases a decrease of the cut-off
radius entails a higher density of the galaxy system. As we can see, if
the galaxy distribution is initially
denser, the system has a faster evolution rate. Since this is also
the effect of a more concentrated distribution, as discussed earlier,
we can conclude that 
it is the density in the central parts that controls the rate of
evolution. Thus distributions with higher central densities, coming
either from higher central concentrations or higher densities over a
larger area, will have a faster evolution than distributions with
lower densities in the central area.
 
Fig.~\ref{evol3} compares the evolution of simulations
P10f1r1 and P10f1r20 and shows
the effect of the ratio of the mass in individual galaxies to the mass in the
common halo. Simu\-lation P10f1r1 has 20 times more mass in the common
halo than simulation P10f1r20. The comparison shows clearly that
the simulation with the lower 
fraction of the mass in galaxies (P10f1r1), i.e. with the higher
fraction of the mass 
in the common halo, evolves slower than the simulation with the higher
fraction of the mass in galaxies.
This in agreement with what was seen in the
simulations of Bode et al. (1994) for poor clusters of galaxies, as
well as those of  
Barnes (1985), Bode et al. (1993) and Athanassoula et al. (1997) for
compact groups. This may be easily understood (Athanassoula et
al. 1997) if we think of the galaxies as a perturbation of the global
common halo potential. If the fraction of the mass in galaxies is
higher, the perturbation is also 
stronger, the galaxies will attract each other more strongly and have
more chances of colliding and therefore of merging.
The opposite will be true if the fraction of the mass in
galaxies is small.
In the limiting case were the galaxies are considered as test
particles in the 
common halo, they will collide only if their orbits accidentally cross.  

Fig.~\ref{evol4} compares the evolution of two simulations, one,
H10f1r1, in which the distribution of the common halo and the galaxies
follows a Hernquist law, and the other, P10f1r1, in which it follows a
Plummer law, the remaining parameters being the same. The difference
is quite striking. We note that the 
evolution is much more rapid for the 
Hernquist model. Thus, at any given time, for the Plummer model there
are more galaxies that have not fallen into the central one than for
the Hernquist model.

\subsection{Global parameters}
\label{glob-par}

\begin{figure}
\includegraphics{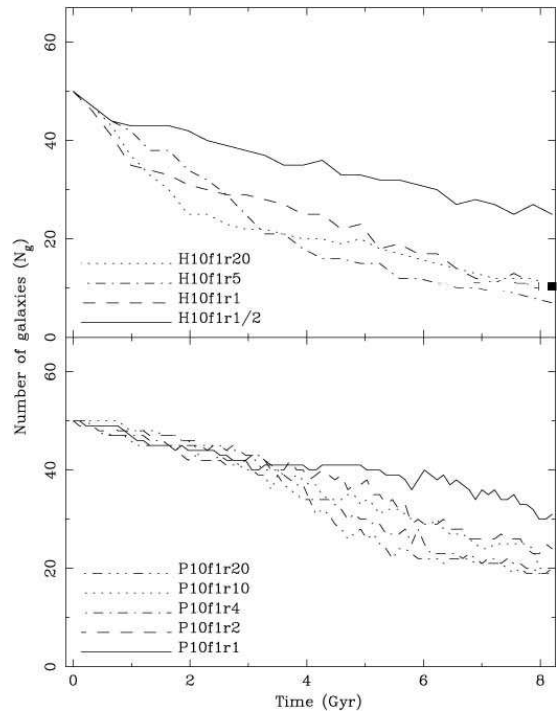}
\vspace{9.6cm}
\caption{Time evolution of the number of surviving satellite 
galaxies. Comparison of the effect of the initial fraction of mass in 
galaxies both for Hernquist (upper panel) and Plummer (lower
panel) mass distributions with the same total energy.
}
\label{evol_number_1}
\end{figure}

\begin{figure}
\includegraphics{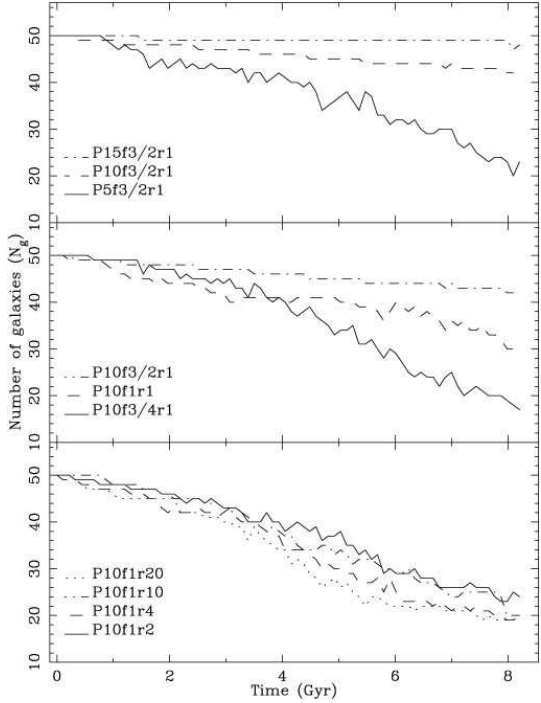}
\vspace{9.6cm}
\caption{Time evolution of the number of surviving satellite 
galaxies. Comparison of the effect of the different parameters of the
initial configuration.
The upper panel shows the effect of the scale-length, 
the middle panel the effect of the cut-off in the
galaxy distribution and the lower panel the effect of the fraction of
mass in galaxies.}
\label{evol_number_2}
\end{figure}

We will use in the analysis of the simulations the same definitions of
the central galaxy and of the
merging conditions as in GAG. They allow us to
make a more quantitative study of the evolution of the 
system. The first
factor that we can consider is the number of galaxies that survive at
a given time. This is plotted as a function of time in
Fi\-gures~\ref{evol_number_1} and \ref{evol_number_2}. In the upper 
panel of Fig.~\ref{evol_number_1} we show the results for the 
simulations with a Hernquist distribution 
and in the lower panel for the simulations with a Plummer distribution. In
this figure we 
concentrate on simulations with $a_p = 10$ and $f=1$ and vary only the
fraction of mass in the galaxies. The evolution shown in
Fig.~\ref{evol_number_1} 
is in good agreement with the qualitative results of
Fig.~\ref{evol3}. In general a larger percentage of the total mass in galaxies
leads to a faster evolution, for the reasons explained in the previous
section. For
simulations with Plummer distributions the effect is only important
in the second half of the evolution when the central galaxy is
sufficiently massive. On the other hand for simulations
with Hernquist distributions the effect is clear from the start, and
is stronger. These differences are due to the different 
evolutionary histories of the two types of simulations, which will be
discussed in 
section~\ref{sec:mass_increase}. Finally we should also note that the
fastest of all evolutions is for
simulations with Hernquist profiles and the biggest fraction of the
mass in galaxies.

Fig.~\ref{evol_number_2} also shows the number of galaxies that have not
merged at a given time as a function of this time, but concentrates on
Plummer profiles, to see the effect of the three free parameters of
the Plummer distribution. The upper panel illustrates the effect 
of the scale-length of the distribution of the common halo and of the
galaxies, the middle one the effect of the cut-off radius
and the lower one the effect of the fraction of mass in the galaxies.
The results agree well with what is seen qualitatively in
Figures~\ref{evol1} to \ref{evol3}. 
The galaxy number decreases faster in the cases of more concentrated mass
distribution, i.e. halos and galaxy distributions with a smaller scale-length.
Also configurations with a smaller cut-off radius of the
galaxy distribution evolve faster.

\begin{figure}
\includegraphics{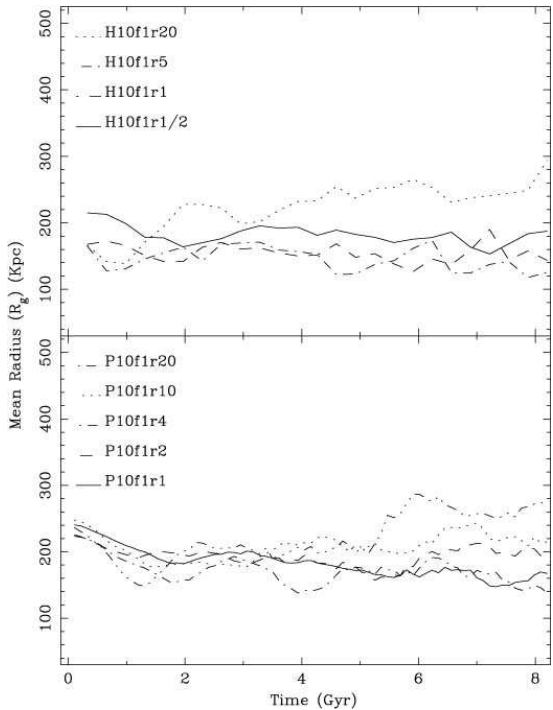}
\vspace{9.6cm}
\caption{Time evolution of the mean radius of the galaxy system of surviving 
galaxies. The layout of the figure is as for Fig~\ref{evol_number_1}.}
\label{evol_radius}
\end{figure}

The next quantity that we can consider is the evolution of the mean radius of 
the system of surviving galaxies. It is calculated as the average
distance of the centers of the surviving galaxies from the center of
the group not taking into account the giant
central galaxy. 
Its time evolution is shown in Fig.~\ref{evol_radius}. As in
Fig.~\ref{evol_number_1} we show in this figure 
the simulations with $a_p=10$ for the Hernquist halo distributions (upper
panel) and Plummer distributions (lower panel). In all but a couple of
cases the mean radius of the system stays constant. This argues that
the satellite 
galaxies that fall into the central one are not only those that have
lower binding energies and orbits with a smaller mean radius. Rather all 
galaxies, independent of their binding energy, are concerned.
The case of H10f1r20 is different, since the mean radius of the
surviving galaxies increases with time. This is due to the fact that
one galaxy is on an outwards going orbit, while the total number of
galaxies is rapidly reduced (cf. Fig.~\ref{evol_number_1}) so that
this one galaxy can influence the statistics.

\begin{figure}
\includegraphics{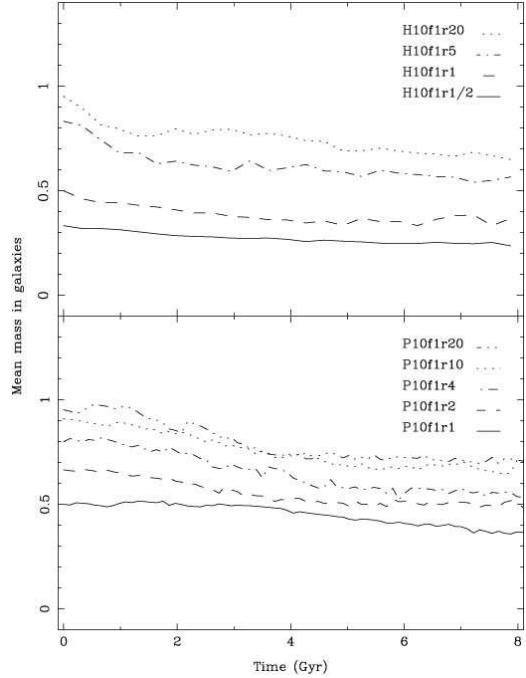}
\vspace{9.6cm}
\caption{Time evolution of the mean mass of the surviving
galaxies in computer units. The layout of the figure is as for
Fig~\ref{evol_number_1}.} 
\label{mass_in_galaxies}
\end{figure}

The mean mass in surviving galaxies, i.e. not taking into account the giant 
central galaxy, is indicative of whether the mass loss
effects due to stripping 
are important in the dynamical evolution of the groups, or not. The
time evolution of 
this parameter is shown in Fig.~\ref{mass_in_galaxies} for 
Hernquist systems (upper panel) and 
for Plummer systems (lower panel). In all the cases the mass in galaxies
diminishes with time, which means that stripping is present and that
the stripped mass is not all captured by other individual
galaxies. Some of it is captured by the giant central galaxy, while
another part stays in between the
individual galaxies.

\subsection{The rigid halo case}
\label{sec:rigid}

If we could, in general, use in simulations a rigid halo, rendered by
an external forcing, constant in time, instead
of a ``live'' halo 
formed by a large number of particles, we could save a considerable 
amount of CPU time. For this reason we will, in this section, compare
the effects of a rigid and a live halo on the evolution of a
simulation, to see whether the corresponding gain in CPU is possible
without introducing a bias in the results. At the beginning of the
simulation a live halo will give, on average, the 
same forces as a rigid halo. This will, however, not be necessarily the
case as the simulation evolves, since the mass distribution of the
halo may change with time, in response to the evolution of the
distribution of the individual galaxies and to the formation of the
giant central galaxy. Furthermore, even in cases where the mass
distribution in the common halo hardly evolves, we are not assured of
the adequacy of the rigid halo, since it 
will not reproduce the dynamical friction on the satellite galaxies that  
the live halo exerts naturally. This effect could in principle be estimated
by the Chandrasekhar (1943) formula, except for the fact that the
common halo does not necessarily have a homogeneous density
distribution, while dynamical friction is much
more important in strongly centrally concentrated backgrounds than in
less concentrated ones (Athanassoula, Makino \& Bosma 1997). It
is thus necessary to compare a simulation with a live halo to one with
a rigid one, to see how big a bias the rigidity of the halo introduces.

\begin{figure}
\includegraphics{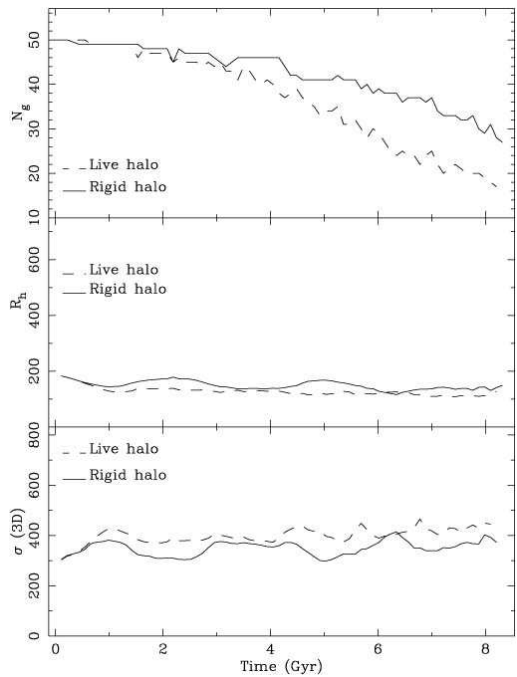}
\vspace{9.6cm}
\caption{Effect of the rigid halo on the evolution of the global
properties of the system. The upper panel compares the number of
surviving galaxies, the middle one their mean radius and
the lower one their velocity dispersion.}
\label{rigido}
\end{figure}

To check this effect we have repeated one simula\-tion, namely P10f3/4r1,
 using
a ri\-gid ha\-lo
ins\-tead of the live halo com\-posed of 50000 parti\-cles. In
Fig.~\ref{rigido} we com\-pare the 
evolu\-tion of the glo\-bal para\-meters of the two systems. 
We can see that the number of surviving galaxies
is considerably lower in the case of a live halo (upper panel). The principal
reason for this is the fact that the dynamical friction with the live halo 
brakes the
galaxies and, as they move slower, they merge easier with the central
galaxy. The radius which
contains half the mass of the galaxy system stays in both cases roughly
constant with time and the average value is roughly the same in the
two simulations (middle panel). 
The velocity dispersion of the system of galaxies shows in both cases
an increase with time, presumably due to the fact that the faster
individual galaxies merge less than the slower one. The lower panel of
Fig.~\ref{rigido} shows that this increase is stronger for the live
halo than for the rigid one. This should be linked to the fact that 
there are more mergings in the live halo case. 

\subsection{The effect of the cutoff on the galaxy distribution.}
\label{cut-off}

\begin{figure}
\includegraphics{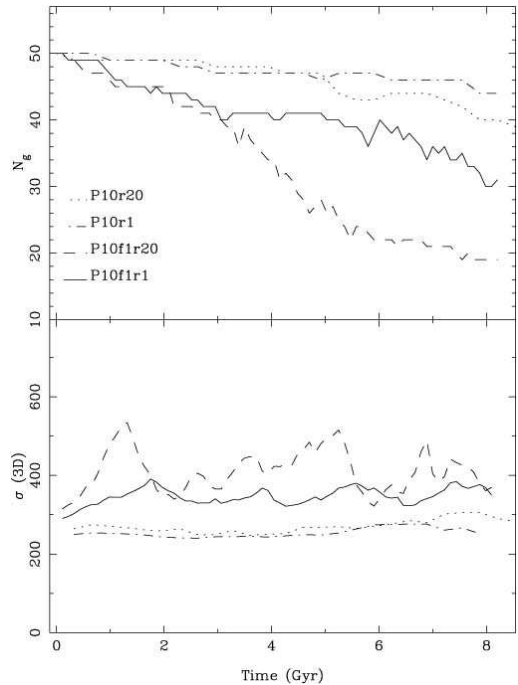}
\vspace{9.6cm}
\caption{Effect of a cutoff in the initial distribution of 
the galaxies on the global properties of the system. The upper panel
compares the number of surviving galaxies and the lower one
their velocity dispersions.}
\label{cut-off-f}
\end{figure}

In most of our simulations we have included an artificial cutoff in
the initial distribution of the galaxies. This partly reflects the
fact that
the distribution of galaxies in a group or cluster also does not
exceed to infinity. It has the added advantage of improving the
statistics for a fixed number of particles per simulation. In this
section we discuss 
its effect by comparing the evolution of runs with cut-off with that
of runs without cut-off and otherwise the same
properties. Fig.~\ref{cut-off-f} compares two such sets of
simulations, simulation P10r20 to P10f1r20, and simulation P10r1 to
P10f1r1. In this figure 
we compare two global properties of the system of surviving galaxies:
their number (upper panel) and their velocity dispersion (lower panel). 
The number of surviving galaxies decreases considerably faster with
time if we include a 
cut-off. This is due to the fact that by constraining the volume,
while keeping the number of galaxies fixed, we increase the density
of the galaxies and this in turn, as discussed in the beginning of
this section, leads to a faster evolution of the system. 

The lower panel shows that the velocity dispersion is smaller for the
cases without cut-off. There can be two possible explanations to
this. The first one relies on the fact that it is the galaxies with
the lower velocities that will merge, and therefore systems that
involve more merging should also have a higher dispersion of
velocities. If, however, this was the explanation then we would note
differences in the velocity dispersions after the differences in the
number of remaining galaxies occurred, which, as can be seen in
Fig.~\ref{cut-off-f} is not the case. We will thus explain the
difference by the fact that the
velocity dispersion of the group decreases with radius. Thus a more
extended group has a lower mean dispersion of velocities than a more
compact one.

\section{The central object}
\label{sec:central_gal}

The central object is defined as in GAG. The properties of
this object are studied using the same methods described in that paper and we
refer the reader to that paper for more details. Here we will discuss
directly the physical properties of the objects formed in our simulations.

\subsection{Stripping versus merging}
\label{sec:mass_increase}

\begin{figure}
\includegraphics{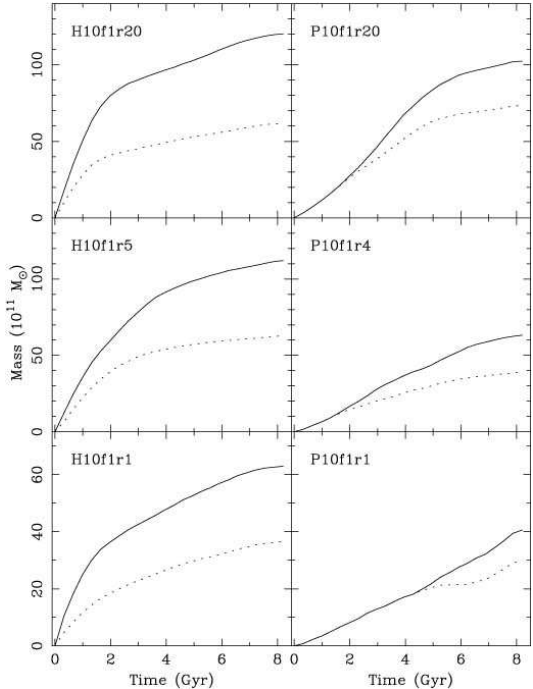}
\vspace{9.5cm}
\caption{Time evolution of the total mass in the central parts, 
excluding the mass in individual galaxies. Left 
panels correspond to  Hernquist simulations and right panels to Plummer
simulations. The solid  
line corresponds to the evolution of the total mass while the dotted line 
gives the fraction of this mass that comes from tidal stripping. The
name of each simulation is given in the upper left corner of each
panel.}
\label{mass_origin}
\end{figure}

The mass of the central galaxy grows with time by means of two mechanisms: 
by merging of satellite galaxies and by accreting material tidally stripped 
from galaxies. We are interested in how the relative importance
of these two  
mechanisms depends on the initial conditions. In
Fig.~\ref{mass_origin} we show for some selected simulations the time  
evolution of the total mass in the central regions not belonging to the
individual galaxies. In the left panels we show the simulations where the
initial density  
distribution follows a Hernquist profile, while in the right panels we show 
the cases of initial Plummer density distributions. Each horizontal
couple of panels corresponds to two simulations with identical or
comparable global properties. The 
solid line corres\-ponds to the total mass as a 
function of time, while the dotted line corresponds to the part of this 
mass that comes from stripped material. The difference between the two lines 
at a given time gives the part of the total mass that comes from mergers
with satellite galaxies. 

Let us first note that in general the total mass of the
central object is larger in simulations starting with Hernquist
profiles than in simulations starting with Plummer profiles. 
The total mass of the central object (at a
given time) is also larger in simulations where the total mass that is
initially in galaxies is larger, as expected.

As can readily be seen from Fig.~\ref{mass_origin} the initial distribution of 
the mass profile (background mass and distribution of galaxies) is
determinant not only for the total mass of the central object, but
also for the formation process leading to it.  
In Plummer simulations all the initial mass increase
is due to the accretion of stripped material. This is true for
a longer period of time 
in the case of simulation P10f1r1, where the mass in galaxies was 
initially the same as the mass in the common background. In this simulation 
also the final fraction of  
the total mass that comes from tidal stripping is more important than
in the others.
On the other hand in the Hernquist simulations the merging of
satellite galaxies into the central galaxy is present 
from the start. This
should be due to the differences in central concentration. Our results
argue that the
central concentration of the Hernquist profile is sufficiently high to
compensate for the higher relative velocities of the galaxies and
to cause a high initial burst of mergings.

Thus, as can be seen from Fig.~\ref{mass_origin} the relative
importance of the stripping is more significant in the 
Plummer cases than in the Hernquist ones. Conversely the absolute
amount of stripped material in the first part of the simulation is
larger for the Hernquist cases. This difference is due to the fact
that the total mass of the central object in the first part of the
simulation is considerably bigger for Hernquist cases. 

These differences between simulations with Plummer and with Hernquist
profiles can be readily understood from the different evolutionary
histories of the two groups of simu\-lations.
In the Plummer case the mass distribution in
the central parts is nearly flat and as the initial
position of the satellite galaxies is selected following this 
distribution, no initial concentration of galaxies is present. Thus, there 
is no strong merging of satellite galaxies into a central object during the 
first
steps of the simulations. The satellite galaxies orbiting within the core radius
occasionally merge with some of their companions. 
These mergings, which do not necessarily occur in the center-most parts 
increase the number of larger galaxies within the core radius, until two or 
more of 
these finally merge to give a giant central object.  

On the other hand, the dynamical 
evolution of the simu\-lations with a Hernquist profile is very
different. The strong peak 
of this density profile towards the central parts 
gives a strong initial concentration of satellite galaxies which merge very 
fast to create the seed of a giant central object. We refer to this stage of the
evolution as an initial burst of mergers. The central object will accrete some  
stripped material and merge with some other satellite galaxies entering the 
central parts. After the strong initial merger, the mass increase of the 
galaxy is milder than in the initial steps of the simulations. As we will 
see in subsection~\ref{sec:profiles} these two different formation stories give galaxies
with very different observational properties. Not all the central objects will
have profiles that resemble real giant elliptical or cD galaxies.

\subsection{Volume density profiles.}
\label{sec:vprofiles}

\begin{figure}
\includegraphics{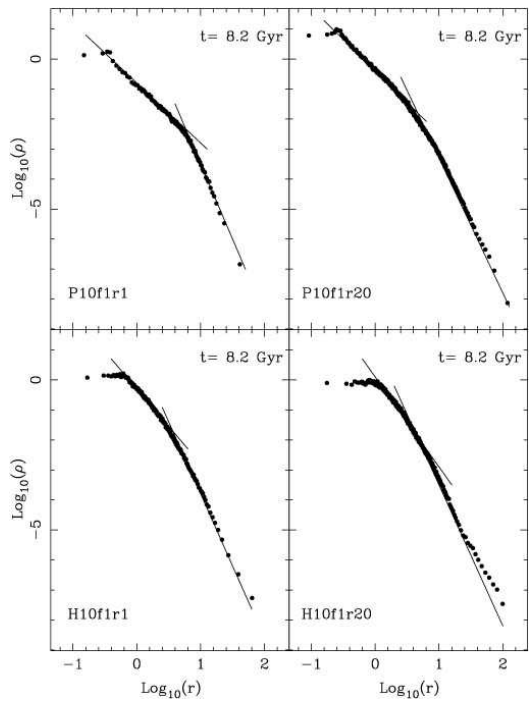}
\vspace{9.5cm}
\caption{
Volume density profile, given by filled circles, as a function of radius
for simulations P10f1r1 
(upper left), P10f1r20 (upper right), H10f1r1 (lower left) and
H10f1r20 (lower right). The time is given in the upper right corner of
each frame. The solid lines give eyeball fits by straight lines. 
}
\label{vprof}
\end{figure}

Figure~\ref{vprof} shows the volume density profile as a function
of radius for simulations P10f1r1, P10f1r20, H10f1r1 and H10f1r20 at 
$t = 8.2$ Gyrs. Plummer simulations show clearly two distinct regions,
the outer one being much steeper than the inner one. Each of these can
be relatively well fitted by a straight line in a  
log-log plane. The slopes of these two lines, as obtained by eyeball
fits, are roughly -2 and -5. The outer part is clearly discernible all
through the simulation and its slope, which is roughly equal to 
the slope of the outer part of the Plummer density profile,
stays roughly constant all through the simulation. This is not the
case for the inner part, which in some earlier times is more difficult to
discern than at $t = 8.2$ Gyrs. Since our simulations stop at $T =
8.4$ Gyrs, we have no information on the evolution of the profile
at later times. 
 
The density profiles coming from Hernquist simulations do not show the
clear existence of two distinct regions, although they do 
steepen at larger radii. For consistency we also made eyeball fits
with two straight lines. The inner slopes thus found are steeper than the
corresponding ones for the Plummer simulations, while the slopes of
the parts further out are roughly the same. The profile of H10f1r20
has a shallower sloper in the outermost parts, which will be discussed
at length in the next subsection.

We compared all these profiles to the Hernquist profile (Hernquist
1990) and found
that, if we exclude the innermost and outermost parts, the Plummer
simulations with little mass in the common halo, as 
well as all Hernquist simulations, are well fitted by this
profile. This was,
however, not the case for the Plummer simulations 
with a high fraction of the mass in the common halo. The physical
implications of this result will be
discussed further in the next subsection with the help of the
projected density profiles.

\subsection{Projected density profiles.}
\label{sec:profiles}

The most interesting property of the central galaxies formed in these
simulations appears when we study their projected density distributions. 
Remnants of the merging of a pair of progenitor galaxies have 
a projected density profile which is well fitted, in all but the innermost
parts, by the $r^{1/4}$  
law (see for instance Barnes and Hernquist 1992, 
or Barnes 1998 and references therein).
A similar result was found for multiple mergings in compact
groups (Weil \& Hernquist 1996, Athanassoula \& Vozikis
1999). Furthermore  
the objects formed in  GAG from multiple 
mergers in large groups or small clusters also showed this property,
except for an excess projected density in the outermost parts in some
cD-like cases.   
Our new simulations with a common halo present a larger variety of profiles. 

\begin{figure}
\includegraphics{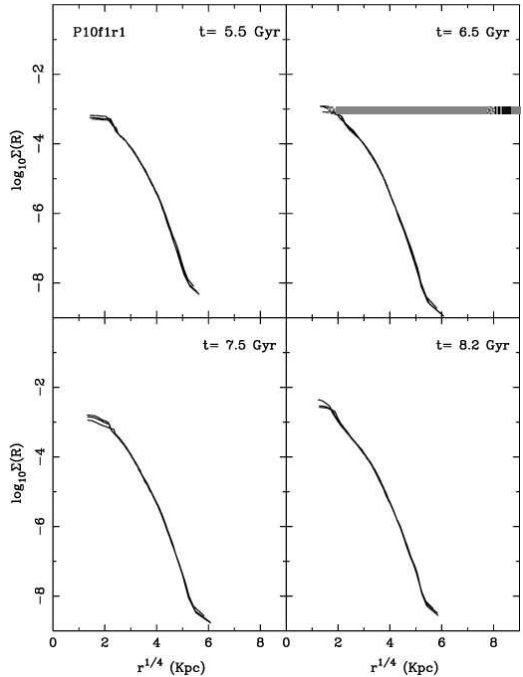}
\vspace{9.2cm}
\caption{Projected density profiles of the central object formed in
si\-mulation P10f1r1 at four different
times. It is clear that the profiles are not well described by $r^{1/4}$
laws. In each panel we give by solid lines the result of 10 random
projections. The time is given in the upper right corner of each
panel and the projected density is in arbitrary units. }
\label{p1}
\end{figure}

\begin{figure}
\includegraphics{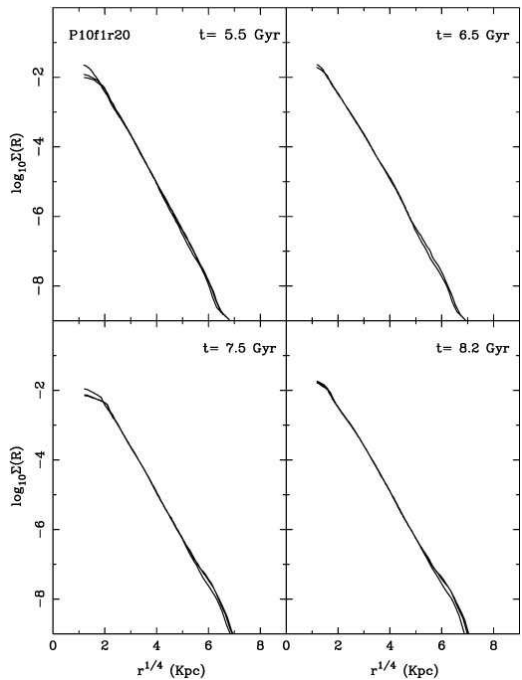}
\vspace{9.2cm}
\caption{Projected density profiles of the central object formed in
simulation P10f1r20 at four different times. It is clear that the
profiles are 
well described by $r^{1/4}$ laws. The layout is as for the previous figure. }
\label{p20}
\end{figure}

The three-dimensional density profiles of the central 
object formed in the Plummer models with a considerable fraction of the mass in
the background halo can not be well fitted by a Hernquist law, and
therefore does not give an $r^{1/4}$ in projection. In
fact the larger the fraction of mass in the common halo, the worst 
the agreement is. On the contrary, if the mass in the common halo is very
much smaller than the mass in common galaxies (e.g. in simulation
P10f1r20), then  
we recover initial conditions closer to those of the simulations in 
GAG, and the profiles are better fitted by a Hernquist law. This form of the
profiles
does not depend either on the initial scale-length of the mass distribution, 
or on the radius initially containing all galaxies, 
and depends only on the initial mass fraction in galaxies. These
results are best illustrated in Figures~\ref{p1} and \ref{p20}.

The mass distribution in the central parts of simulation P10f1r1, where
matter is equally distributed between the galaxies and the background,
is heavily influenced by the matter in the common halo, and the
violent relaxation which would have otherwise led to an $r^{1/4}$
projected profile is hampered.
This result can be seen in Fig.~\ref{p1}, where we show
the projected 
density profile of the central object formed in simulation P10f1r1 at four
different times. In each of the panels we show the result of 10
projections from random viewing angles, thus making clear that this
result is projection  
independent. At no time does this profile follow an $r^{1/4}$
law and thus the central object cannot be associated to any giant
elliptical or cD 
galaxy, as are observed in the central parts of galaxy clusters.

The central object formed in simulation P10f1r20 feels only a small 
influence from the common halo and develops a density
distribution  
which is well fitted by an $r^{1/4}$ law. This can be seen in
Fig.~\ref{p20}, where 
we show the projected density profile of the central object at four
different steps of the simulation. This effect is again not due to a
particular viewing angle. As was already mentioned, this result
is very similar to what was found in GAG since in P10f1r20 there is
very little influence from the background.  

\begin{figure}
\includegraphics{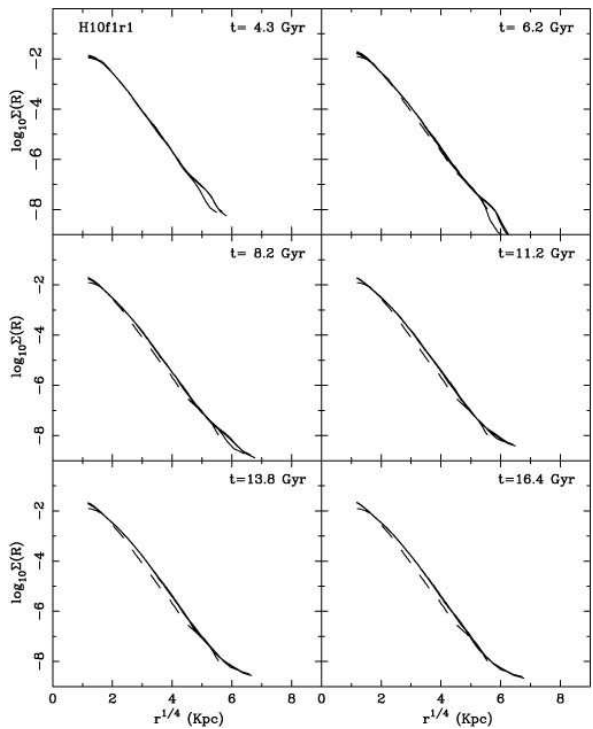}
\vspace{10cm}
\caption{Projected density profiles of the central galaxy formed in
simulation H10f1r1 at six different
time steps. The profiles are well described by $r^{1/4}$ laws in the
main body of the object and show luminosity excesses over the
$r^{1/4}$ in the external parts. In each panel we give by solid lines
result of 10 random 
projections and by a dashed line we repeat one of the profiles of the
upper left panel. The time is given in the upper right corner of each
panel and the projected density is in arbitrary units. }
\label{h1} 
\end{figure}

Simulations where the common background is described initially
by a Hernquist distribution give very different results. The projected
density profile for the 
central galaxy in simulation H10f1r1 is shown in Fig.~\ref{h1}. As in the case 
of simulations P10f1r1, only half the mass was 
initially in the galaxies.  
We note that the
projected density profile of the central object follows 
an $r^{1/4}$ law in the main body of the object and 
can thus be associated to a real giant elliptical. 

\begin{figure}
\includegraphics{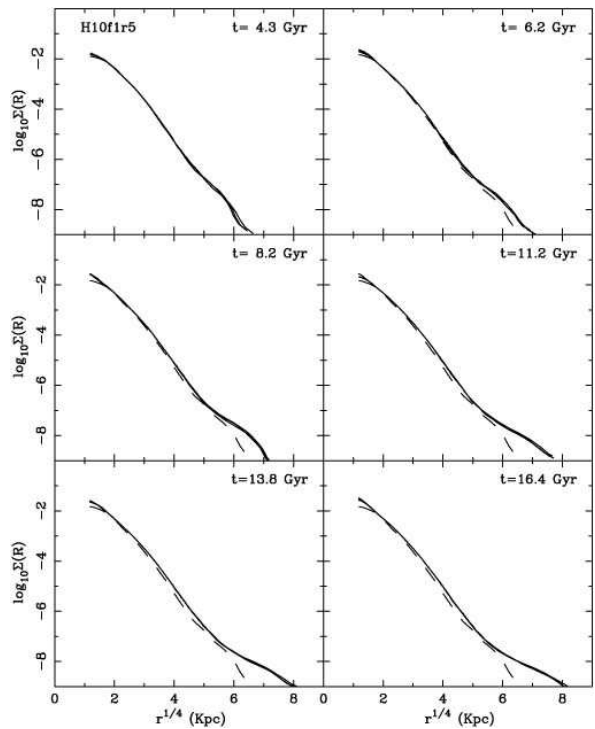}
\vspace{10cm}
\caption{Same as the previous figure but for simulation H10f1r5.
}
\label{h5} 
\end{figure}

\begin{figure}
\includegraphics{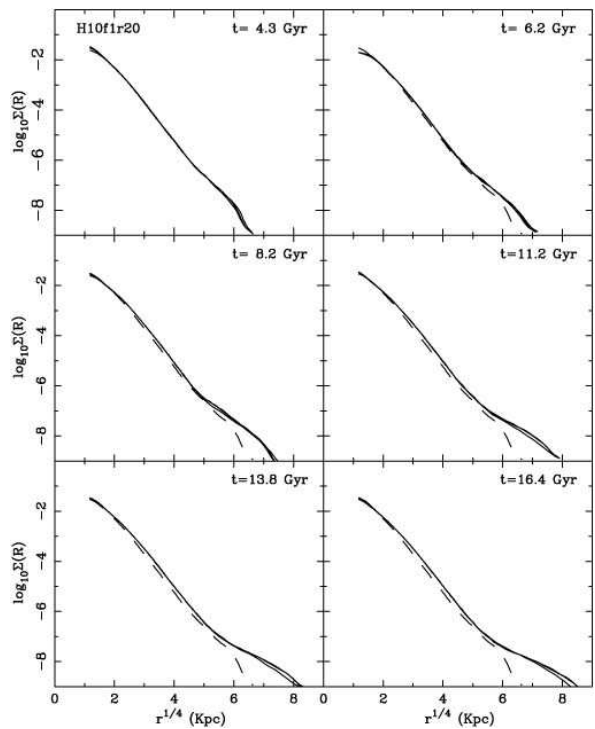}
\vspace{10cm}
\caption{Same as in figure \ref{h1} but for simulation H10f1r20. }
\label{h20}
\end{figure}

Simulations H10f1r5 and H10f1r20 have a much smaller fraction of their
mass in the common halo; roughly 17 and 5 per cent respectively.
Their projected radial density profiles are shown for six different
time steps in Figures~\ref{h5} and \ref{h20}. They show clearly that
the radial profiles follow well the $r^{1/4}$ law over the main body of the
object. Here it can not be claimed that this profile is formed as a
response to the massive common halo with a Hernquist radial profile,
since the mass in the halo is rather small. These central objects were
formed in a situation where there is a strong central concentration of
galaxies at the start, thus favouring a strong and rapid merger. Thus giant
galaxies formed from a rapid multiple merger in the central part of
groups or clusters should have an $r^{1/4}$ projected density profile
over their main body.

Now let us turn to the excess mass over the $r^{1/4}$ law in the
outermost parts of the 
central object. This should be associated with the corresponding
excess light in the 
outermost parts of some BCM's and giant elliptical
galaxies. This light excess has, unfortunately, been called ``halo'', a
term which could bring confusion with the dark halo. To avoid such a
confusion, while being coherent with the nomenclature used so far, we will use
the term ``luminous halo''. Giant ellipticals with such a luminous halo are
called cD galaxies. We will now examine which of our simulations
present central objects with such a luminous halo and which can therefore be 
associated with cD galaxies. 

Simulation P10f1r20, which is the only one of those with an initial
Plummer mass distribution that shows an $r^{1/4}$ profile over the
main body of the galaxy, has only a very small density excess over the
$r^{1/4}$ profile in the outermost parts of the central
object, which hardly grows with time. It can thus not be easily
associated with a cD galaxy. 

Central objects formed in simulations with a Hernquist background 
present such an excess of matter over the $r^{1/4}$ profile in the
outermost parts. This is small in cases with a large fraction
of the mass in the common background, as H10f1r1, and becomes more
important for 
simulations where the fraction of mass in the galaxies is also more
important. Thus the central object in simulation H10f1r5 has a clear
outer mass excess and that in H10f1r20 an even bigger one
(cf. Figures~~\ref{h5} and \ref{h20}),
independent of 
the projection selected. The existence of this extra mass is clear
relatively early 
on in the simulations, and increases considerably with time. Thus such
objects should be associated to cD galaxies, since they form in the
central part of the group, follow an $r^{1/4}$ projected density law
over the main part part of their body and have excess matter over that
profile in the outermost parts.

This inverse correlation between the excess mass in the outermost parts and the
amount of mass in the common background halo could be due to
the fact that a deviation from the $r^{1/4}$ is more difficult to form
in the strong $r^{1/4}$ background. Alternatively it could be due to
the fact that the rate of evolution is slower in simulations with a
massive common halo, i.e. that at a given time the three simulations
are at different stages of their evolution. The total evolution time in these
simulations was twice as long as that of Plummer simulations
in order to study the possible influence of the dynamical
evolution of the group in the distribution of the luminous 
halo material of the central
object. In Figs.~\ref{h1}, \ref{h5} and \ref{h20} we show that indeed the 
luminous
halo of the central object increases with time, specially in the cases where
there is a considerable fraction of mass initially in the galaxies. 
As we have already pointed out in section 2, the conversion from computer units
to physical units is not unique and thus the times given in figures \ref{h1},
\ref{h5} and \ref{h20} could be modified if e.g. a different value was adopted
for the masses or the scale-length of the initial galaxies. We nevertheless 
believe that
all simulations were evolved sufficiently long for a clear picture to be
obtained.

\section{Limitations of our models}
\label{sec:assume}

Simulations of the evolution of galaxy clusters, seen the complexity
of the objects to be modelled, use a number of
simplifications and assumptions, and the simulations presented here
are no exception. We will here discuss the simplifications and
assumptions specific to our simulations to get a better understanding
of how they might have influenced our results.

The initial number of galaxies in all our simulations is 50. Thus
they do not apply to rich clusters, but more to large groups, or poor
clusters. They can also apply to sub-condensations in rich clusters 
-- provided of course the
remaining part of the cluster does not significantly influence the
evolution -- or subunits which will come together 
to form a large cluster. 

Our initial conditions are not cosmological. They were constructed
specifically 
so as to allow us to study best the physics of the formation of the
central object and particularly the effect of different physical parameters
on this formation. We have thus created series of initial conditions
where one parameter only was varied, so as to best put forward the
effect of this specific parameter. We considered in turn the effect of the
scale-length of the mass distribution, of the cut-off radius of the galaxy
distribution and of the mass fraction in the common halo. Most
important, 
we have considered both matter distributions with a core and matter
distributions with a 
cusp, and have studied the effects of these two very different
distributions on the evolution of the group and on 
the properties of the central galaxy. In other words we have been able to create our initial
conditions such as to put forward best the phenomena we wished to
study. Of course cosmological simu\-lations and in particular
observations will be 
necessary to tell us whether all the spectrum of initial conditions
considered here are possible or realistic. Our approach is similar to that
used e.g. in studying the formation of bars in disc galaxies. It is 
possible to conclude that the growth rate of the bar is smaller in
discs which are hot and/or have an important spheroidal component in
their center or main parts, without having to worry about the exact
form of the radial density profile of the disc or halo component. Once
the essential physics of the bar formation is 
understood, then one can start worrying about whether this or that
type of
radial density profile is more realistic, follows best from
cosmological initial conditions, or is in better agreement with the
observations. 

The fraction of mass in the common halo is, in all our simulations,
considerably 
smaller than what is believed to be the case in real groups or clusters.
Nevertheless, the range explored is sufficient for us to establish how
the fraction of mass in 
the common halo influences the evolution. It also allowed us to
establish that the form of 
the radial profile of the central object in Plummer simulations is
incompatible with observed profiles in cases with a high fraction of
mass in a common halo. Simulations with a higher fraction of mass in the common
halo would necessitate considerably more CPU time for two
reasons. First the total
number of particles in the simulations would increase considerably if the
number of particles per galaxy is kept constant. Also the time it
would take for 
the central object to form would also increase, thus leading to longer CPU
times, if the time-step is kept constant. Nevertheless, it would be
interesting to pursue  in the future simulation with higher percentage
of mass in the common halo, particularly in order to test the
formation of the luminous halo in the outer parts of the central objects.

Probably the most serious drawback of our simulations is the fact that
our galaxy models are very simplistic. Thus we do not distinguish
between luminous and dark matter, and have not considered any galaxies
with a disc, with anisotropy or with rotation. Thus we can not distinguish
between dark and luminous mass in the central object. Simulations with
more realistic galaxy models 
would necessitate a considerably higher number of particles,
particularly if disc galaxies are considered. Such simulations would
ne\-vertheless be very interesting, and should be envisaged now that
increases in computing power make them possible.

We have also neglected the existence of gas, both cold and hot, both
in the galaxies and in the common background halo. Thus effects like
ram-pressure stripping of individual galaxies are totally neglected,
but this should not influence the formation of the central object. It
should of course be kept in mind that the fraction of common
background halo in our simulations includes all mass encompassing the
whole cluster and not linked to individual galaxies, i.e. it includes also
the hot gaseous component.
 
The above summarise the main simplifications and assumptions underlying
our simulations. It is obvious that there is room for improvement, but
none of our assumptions should alter the main physical results found
in this paper.

\section{Summary and Discussion}
\label{sec:summary}

We have performed N-body simulations of the dynamical evolution of groups of
50 galaxies embedded in a common background mass distribution. We
considered two density distributions, the
Plummer and the Hernquist one. The former
has a core, while the latter has a cusp. The main goal of this 
paper was to compare the evolution of the group and the properties of
the central giant galaxies formed in these 
two very different matter distributions. 

As in the case of simulations of groups with no common background
distribution of matter (GAG), the evolution of the systems within a common
halo is driven mainly by the merging instability (Carnevali et al. 1981).
In all cases a giant galaxy is formed in the central parts by merging of
some secondary galaxies and by accreting material stripped from the rest of the
galaxies by tidal forces. The evolution rate of the systems is very
sensitive to 
the initial conditions and is controlled mainly by the density in the
central parts. Distributions with higher densities resulting either from
higher central concentrations or higher densities over a larger area, have
a faster evolution than distributions with lower densities in the
central parts. One should nevertheless remember that the dynamical time of
denser configurations is shorter.

To study more quantitatively the evolution of the different simulations we have
followed the time evolution of some global parameters, like the number
of galaxies that have not yet merged, their mean distance from the
center of the group, their mean mass and their velocity
dispersion. The total number of 
galaxies decreases with time in all cases. This decrease is faster for
configurations with a larger fraction of the total mass in galaxies.
This temporal decrease is particularly important for the first $2$ Gyrs 
of the evolution of cusped initial
distributions, where there is a strong initial merging. In the Plummer
simulations, 
where the initial dominant effect is the tidal stripping of material, 
the temporal decrease of the number of galaxies is only important 
after the first $2$ Gyrs. The
effect of the initial conditions on the mean distance of the galaxies from the
center of the group, on their mean mass and on the velocity
dispersion of the galaxy system is less spectacular.

We have also studied the effect of substituting the live common
background halo of particles by a 
rigid halo with the same parameters by comparing two simulations with
the same initial conditions, except that the one has a rigid and the
other a live halo. The global evolution of the two simulations is
quite different. In particular we find that the number of galaxies
that have not yet merged decreases much faster in the simulation with the
live halo. This can be easily understood since the
evolution in a live halo includes the effects of dynamical
friction which brakes the 
galaxy motion and favours the merging with the giant galaxy.
Thus, although simulations with rigid common halos require
considerably less CPU, they should be avoided, since their results are
not reliable.

The different profiles of the mass distribution influence both the
evolution of the group and the properties of the giant galaxy formed
in the centre. In the
case of a Plummer density distribution the individual galaxies undergo
pairwise mergers giving rise to some
larger galaxies in the core of the cluster. At
some later step these 
galaxies merge and form the giant central galaxy. On the other hand,
in the Hernquist case the high central concentration of galaxies in
the center leads to a high ```burst'' of mergers in the central
parts during the initial steps of the evolution and the giant
central galaxy is quickly formed.

Not only the merging histories but also the quantity of stripped
material from the satellite galaxies is different in the two cases.
In the Plummer cases the fraction of the total mass in the central
galaxy that comes from stripping is very important, especially early on
in the simulation. In the Hernquist cases the
fraction of the 
mass of the central galaxy that comes from stripped material is not so
important, but amounts nevertheless to a considerable mass, since the
total mass in the central object is important.

These different evolutionary histories give rise to central objects with very
different observable properties. We have studied the projected density
distributions of the giant galaxies at different time-steps and for
different random projections. The shape of these profiles depends heavily on
the initial distribution of the mass. In the cases of
Plummer distributions with half the total mass in the common 
background, the projected density profile of the giant galaxy does not
follow the $r^{1/4}$ law at any time during the simulation. The 
radial density profiles of these central objects 
are not observed in the real giant galaxies in the center of the
clusters. Only for the case where most of the mass is initially in the
galaxies does the giant galaxy formed within a Plummer distribution follow the
$r^{1/4}$ law typical of elliptical galaxies. 

On the contrary, the galaxies
formed within Hernquist distributions do follow $r^{1/4}$ laws in the main
parts of their bodies
in all the cases. Moreover, in the external parts, they show the typical
projected density excess of cD galaxies in the case where a
considerable fraction of the mass is 
initially in the galaxies. This excess over the $r^{1/4}$ law is not a
projection 
effect and thus reflects the real matter distribution of the central object.
Furthermore, this excess gets larger as the
dynamical evolution 
of the group proceeds. 

Thus our simulations suggest that central galaxies with $r^{1/4}$
profiles can form either if the fraction of mass in the common halo is
very small, or if the density distribution is cusped. Since the amount
of matter observed in the common background is considerably larger
than what would lead to $r^{1/4}$ profiles for all density
distributions, our simulations argue 
strongly in favour of cusped matter distributions,
at least for those groups and clusters where a BCM has formed. 

\parindent=0pt
\def\rr{\par\noindent\parshape=2 0cm 8cm 1cm 7cm}
\vskip 0.7cm plus .5cm minus .5cm

{\Large \bf Acknowledgments.} We would like to thank the referee, Yoko
Funato, for helpful comments, Albert Bosma for
many useful discussions and Jean-Charles Lambert for his 
help with the administration of the runs.
E.A. would also like to thank IGRAP, the
INSU/CNRS and the University of Aix-Marseille I for funds to develop
the computing facilities used for the calculations in this paper.
A.G. and C.G.G. acknowledge financial support by the Direcci\'on de
Investigaci\'on Cient\'{\i}fica y T\'ecnica under contract PB97-0411.
\vskip 0.5cm

%\newpage

{\Large \bf References.}
\vskip 0.5cm
\parskip=0pt
\rr{Aarseth S.J., Henon M., Wielen R., 1974, A\&A, 37, 183}
\rr{Athanassoula E., Bosma A., Lambert J.C., Makino J., 1998, MNRAS,
293, 369}
\rr{Athanassoula E., Makino J., Bosma A., 1997, MNRAS, 286, 825}
\rr{Athanassoula E., Vozikis Ch. L., 1999, in ``Galaxy interactions at low
    and high redshift.'', eds. J. Barnes \& D. Sanders, Kluwer Academic 
    Publishers, Dordrecht, p. 145}
\rr{Bahcall N.A., Fan X., Chen R., 1997 ApJ, 485, L53}
\rr{Barnes J., 1985, MNRAS, 215, 517}
\rr{Barnes J., 1998 in ``Galaxies: Interactions and induced star formation.''
eds. D. Friedli, L. Martinet \& D. Pfenniger, Springer-Verlag, Berlin, p. 275}
\rr{Barnes J. Hernquist L., 1992, ARAA, 30, 705}
\rr{Barnes J., Hut P., 1986, Nat, 324, 446}
\rr{Bode P.W., Berrington R.C., Cohn H.N., Lugger Ph.M., 1994, ApJ, 433, 479}
\rr{Bode P.W., Cohn H.N., Lugger Ph.M., 1993, ApJ, 416, 17}
\rr{Carlberg R.E., Yee H.K., Ellingson E., Morris S.L., Abraham R., Gravel P., 
  Pritchet C.J., Smecker-Hane T., Hartwick F.D.A., Hesser J.E., Hutchings J.B.,
  Oke J.B., 1997, ApJ, 485, L13}
\rr{Carnevali P., Cavaliere A., Santangelo P., 1981, ApJ, 249, 449}
\rr{Chandrasekhar S., 1943, ApJ, 97, 255}
\rr{Cirimelle G., Nesci R., Trevese D., 1997, ApJ, 475, 11}
\rr{Cole S., Lacey C., 1996, MNRAS, 281, 716}
\rr{Cowie L.L., Binney J., 1977, ApJ, 215, 723}
\rr{Dubinski j., 1998, ApJ, 502, 141}
\rr{Eke V.R., Navarro J.F., Frenk C.S., 1998, ApJ, 503, 569}
\rr{Fabian A.C., Nulsen P.E.J., 1977, MNRAS, 180, 479}
\rr{Fort B., Mellier Y., 1994, A\&AR, 5, 239}
\rr{Fukushige T., Makino J., 1997, ApJ, 477, L9}
\rr{Funato Y., Makino J., Ebisuzaki T., 1993, PASJ, 45, 289}
\rr{Gallagher J.S., Ostriker J.P., 1972, AJ, 77, 288}
\rr{Garijo A., Athanassoula E., Garc\'{\i}a-G\'omez C., 1997, A\&A, 327,
930 (GAG)}
\rr{Hausman M.A., Ostriker J.P., 1978, ApJ, 224, 320}
\rr{Henry J.P., 1997, ApJ, 489, L1}
\rr{Hernquist L., 1990, ApJ, 356, 359}
\rr{Jing Y.P., 1999, ApJ, L69}
\rr{Kent S.M., Gunn J.E., 1982, AJ, 87, 945}
\rr{Loeb A., Mao S., 1994, ApJ, 435, L109}
\rr{Makino N., Asano K., 1999, ApJ, 512, 9}
\rr{Markevitch M., Vikhilinin A., Forman W.R., Sarazin C.L., 1999, ApJ, 545,
553}
\rr{McGlynn T.A., Ostriker J.P., 1980, ApJ, 241, 915}
\rr{Mellier Y., 1999, ARAA, 37, 127}
\rr{Merritt D., 1984, ApJ, 276, 26}
\rr{Mirald\'a-Escud\'e J., Babul A., 1995, ApJ, 449, 18}
\rr{Moore B., Gelato S., Jenkins A., Pearce F.R., Quilis V., 2000, ApJ, 
535, L21}
\rr{Moore B., Governato F., Quinn T., Stadel J., Lake G., 1998, ApJ, 499, L5}
\rr{Mushotzky R.F., Scharf C.A., 1997, ApJ, 482, L13}
\rr{Navarro J.F., Frenk C.S., White S.D.M., 1996, ApJ, 462, 563}
\rr{Navarro J.F., Frenk C.S., White S.D.M., 1997, ApJ, 490, 493}
\rr{Ostriker J.P., Hausman M.A., 1977, ApJ, 217, L125}
\rr{Ostriker J.P., Tremaine S.D., 1975, ApJ, 202, L113}
\rr{Richstone D.O., 1976, ApJ, 204, 642}
\rr{Rosati P., Ceca R.D., Norman C., Giacconi R., 1998, ApJ, 492, L21}
\rr{Sensui T., Funato Y., Makino J., 1999, PASJ, 51, 1}
\rr{Tyson J.A., Fischer Ph., 1995, ApJ, 446, L55}
\rr{Vikhlinin A., McNamara B.R., Forman W., Jones C., Quintana H.,
Hornstrup A., 1998, ApJ, 498, L21}
\rr{Weil M.L., Hernquist L., 1996, ApJ, 460, 101}
\rr{Wu X.P., Chiueh T., Fang L.Z., Xue Y.J., 1998, MNRAS, 301, 861}
\rr{Yoshida N., Springel V., White S.D.M., Tormen G., 2000, ApJ, 535, L103}

\label{lastpage} 

\end{document}